\documentclass[12pt]{article}
\usepackage{amsmath,amsthm,amsfonts,amscd,amssymb,eucal,bbm}

\def\RR{{\mathbb R}}
\def\CC{{\mathbb C}}
\def\NN{{\mathbb N}}

\def\diff{{{\rm Diff}^+(S^1)}}
\def\mob{{\rm M\ddot{o}b}}
\def\A{{\mathcal A}}

\def\H{{\mathcal H}}
\def\I{{\mathcal I}}

\def\M{{\mathcal M}}

\def\S{{\mathcal S}}

\def\U{{\mathcal U}}

\def\Z{{\mathcal Z}}

\newtheorem{theorem}{Theorem}[section]
\newtheorem{definition}[theorem]{Definition}
\newtheorem{corollary}[theorem]{Corollary}
\newtheorem{proposition}[theorem]{Proposition}
\newtheorem{lemma}[theorem]{Lemma}

\begin{document}

\title{{\huge {\bf Conformal Covariance and 
Positivity of Energy in Charged Sectors}}}

\author{MIH\'ALY WEINER 
\\ Dipartimento di Matematica 
\\ Universit\`a di Roma ``Tor Vergata'' 
\\ Via della Ricerca Scientifica, 1, I-00133, Roma, ITALY
\\ E-mail: weiner@mat.uniroma2.it}
\date{} 

\maketitle 
\renewcommand{\sectionmark}[1]{} 

\begin{abstract}
It has been recently noted that the diffeomorphism 
covariance of a Chiral Conformal QFT in the vacuum sector 
automatically ensures M\"obius covariance in all charged sectors. 
In this article it is shown that the diffeomorphism covariance and
the positivity of the energy in the vacuum sector even 
ensure the positivity of the energy in the charged sectors.

The main observation of this paper is that the positivity of the 
energy --- at least in case of a Chiral Conformal QFT ---  is a local
concept: it is related to the fact that the energy density, when smeared
with some local nonnegative test functions, remains bounded from below
(with the bound depending on the test function). 

The presented proof relies in an essential way on recently 
developed methods concerning the smearing of the stress-energy
tensor on nonsmooth functions.
\end{abstract}

\section{Introduction}

The positivity of the energy is one of the most important selection 
criteria for a model to be ``physical''. In almost all treatments 
of Quantum Field Theory, it appears as one of the 
fundamental axioms. In the vacuum sector it is usually formulated by 
requiring the positivity of the selfadjoint generator of 
every one-parameter group of future-like spacetime translations
in the representation corresponding to the model.

As an axiom, one may say that it is ``automatically'' true, 
but in a concrete model it is something to be checked. 
In particular, to see what are the charged sectors 
with positive energy for a model given in the vacuum sector
may be difficult (as calculating the charged sectors can 
already be a hard problem).

The present paper concerns chiral components of 2-dimensional
Conformal QFTs in the setting of Algebraic Quantum
Field Theory (see the book \cite{Haag} of Haag). In this 
framework a Chiral Conformal QFT is commonly described by means of 
a M\"obius covariant local net of von Neumann algebras on $S^{1}$. 
The net is said to be conformal (or diffeomorphism) 
covariant if the M\"obius symmetry of the net extends to
a full diffeomorphism symmetry (see the next section for 
precise definitions). Charged sectors are described as 
irreducible representations of the net; their general theory was 
developed by Doplicher, Haag and Roberts \cite{DHR1,DHR2}. 
They proved, among many other things, that if a covariant 
sector has a finite statistical dimension then it is 
automatically of positive energy. In \cite{GL} Guido and Longo 
showed that under some regularity condition the finiteness of the 
statistics even implies the covariance property of the sector.

In the particular case of chiral theories, there are many beautiful 
known relations concerning charged sectors; e.g.\! the 
formula \cite[Theorem 33]{KLM} of Kawahigashi, Longo e M\"uger, 
linking the statistical dimensions of the sectors to the so-called 
$\mu$-index of the net. In relation with the positivity of
energy we may say that the case of finite statistics is more 
or less completely understood \cite{GL,GuLo96,BCL}.
In particular, taking in account the above mentioned formula,
if a theory (in its vacuum sector) is split, conformal and
has a finite $\mu$-index --- which means that it is
{\it completely rational} cf.\! \cite{KLM,LoXu} --- then every
sector of it is automatically of finite statistics and
covariant under a positive energy representation of the
M\"obius group. Yet, although these conditions cover many 
of the interesting cases (for example all SU$(N)_k$ models 
\cite{Xu} and all models with central charge $c<1$, see \cite{KL}), 
there are interesting (not pathological!) models in which it does not 
hold and, what is more important in this context, indeed possessing 
sectors with infinite statistical dimension (and yet with positivity
of energy). This is clearly in contrast with the experience coming 
from massive QFTs (by a theorem of Buchholz and Fredenhagen \cite{BuFre}, 
a massive sector with positive energy is always localizable in a 
spacelike cone and has finite statistics).

The first example of a sector with infinite statistics was constructed by 
Fredenhagen \cite{Fre}. Rehren gave arguments \cite{Re} that even
the Virasoro model, which is in some sense the most natural model, 
should admit sectors with infinite statistical dimensions when its 
central charge $c\geq 1$ and that in fact in this case ``most'' of its 
sectors should be of infinite statistics. This was then actually proved 
\cite{Carpi1} by Carpi first for the case $c=1$ and then \cite{Carpi2} for 
many other values of the central charge, leaving open the question
only for some values of $c$ between $1$ and $2$. Moreover Longo e Xu 
proved \cite{LoXu} that if $\A$ is a split conformal net with $\mu = 
\infty$ then $(\A \otimes \A)^{\rm flip}$ has at least one sector with 
infinite statistical dimension, showing that the case of infinite 
statistics is indeed quite general.

Recently D'Antoni, Fredenhagen and K\"oster published a letter 
\cite{D'AnFreKos} with a proof that diffeomorphism covariance 
itself (in the vacuum sector) is already enough to ensure M\"obius 
covariance in every (not necessary irreducible) representation: 
there always exists a unique (projective, strongly continuous) inner 
implementation of the M\"obius symmetry. (In Prop. \ref{n-cover} we shall
generalize this statement to the {\it $n$-M\"obius} group, which is a 
natural realization of the $n^{\rm th}$ cover of the M\"obius group in 
the group of all diffeomorphisms.) Thus the concept of the conformal 
energy, as the selfadjoint generator of rotations in a given charged 
sector, is at least well-defined. (Without the assumption of 
diffeomorphism covariance it is in general not true: there are M\"obius 
covariant nets --- see the examples in \cite{GLW,koester03a} --- possessing 
charged sectors in which the M\"obius symmetry is not even implementable.) 
What remained an open question until now, whether this energy is 
automatically positive or not. The present article shall settle this 
question by providing a proof for the positivity (Theorem \ref{mainresult}). 

The idea behind the proof is simple. The total conformal energy $L_0$ is 
the integral of the energy-density; i.e.\! the stress-energy tensor $T$ 
evaluated on the constant $1$ function. So if we take a finite partition 
of the unity $\{f_n\}_{n=1}^N$ on the circle, we may write $T(1)$ as the 
sum $\sum T(f_n)$ where each element is {\it local}. Thus each term in 
itself (although not bounded) can be considered in a given charged 
sector. Moreover, it has been recently proved by Fewster and Holland
\cite{FewHoll} that the stress-energy tensor evaluated on a nonnegative 
function is bounded from below. These operators then, being local elements, 
remain bounded from below also in the charged sector. So their sum in the
charged sector, which we may expect to be the generator of rotations in 
that sector, should still be bounded from below.

There are several problems with this idea. For example, as the 
supports of the functions $\{f_n\}$ must unavoidably ``overlap'',
the corresponding operators will in general not commute. To deal with
sums of non-commuting unbounded operators is not easy. In particular,
while in the vacuum representation --- due to the well known energy 
bounds --- we have the natural common domain of the finite energy 
vectors, in a charged sector (unless we assume positivity of energy, 
which is exactly what we want to prove) we have no such domain.

To overcome the difficulties we shall modify this idea in 
two points. First of all, instead of $L_0=T(1)$, that 
is, the generator of the rotations, we may work with the 
generator of the translations --- the positivity of any of them 
implies the positivity of the other one. In fact we shall go one 
step further by replacing the generator of translations with the 
generator of $2${\it-translations}. (This is why, as it has been 
already mentioned, we shall consider the $n$-M\"obius group; 
particularly in the case $n=2$.)
This has the advantage that the function representing the 
corresponding vector field can be written as $f_1+f_2$, 
where the two local nonnegative functions $f_1,f_2$ {\it do not} 
``overlap''. These functions, at the endpoint of their support 
are not smooth (such decomposition is not possible with smooth 
functions); they are only once differentiable. 
However, as it was recently proved \cite{CaWe} by the present author 
and Carpi, the stress-energy tensor can be evaluated even on 
nonsmooth functions, given that they are ``sufficiently regular'', 
which is exactly the case of $f_1$ and $f_2$ (see Lemma 
\ref{finite1.5norm} and the argument before Prop.\! \ref{affiliation}). 
As they are nonnegative but not smooth, to conclude that $T(f_1)$ and
$T(f_2)$ are bounded from below we cannot use the result 
stated in \cite{FewHoll}. However, it turns out to be 
(Prop.\! \ref{affiliation}) a rather direct and simple consequence of 
the construction, thus it will be deduced independently from the 
mentioned result, of which we shall make no explicit use. In fact the 
author considered this construction as an argument indicating that 
if $f \geq 0$ then $T(f)$ is bounded from below (which by now is of 
course proven, as it was already mentioned, in \cite{FewHoll}); see 
more on this in this paper at the remark after Prop \ref{affiliation} 
and in the mentioned article of Fewster and Holland at the footnote in the
proof of \cite[Theorem 4.1]{FewHoll}.

Before we shall proceed to the proof, in the next section
we briefly recall some definitions and basic facts 
regarding local nets of von Neumann algebras on the circle.

\section{Preliminaries}

\subsection{M\"obius covariant nets and their representations}
Let $\I$ be the set of open, nonempty and nondense arcs
(intervals) of the unit circle $S^1 =\{z\in \CC : \,|z|=1 \}$.
A {\bf M\"obius covariant net on $S^1$} is a map $\A$
which assigns to every $I \in \I$ a von Neumann algebra 
$\A(I)$ acting on a fixed complex, Hilbert space $\H_\A$ 
(``the vacuum Hilbert space of the theory''), together with 
a given strongly continuous representation $U$ of $\mob \simeq 
{\rm PSL}(2,\RR)$, the group of M\"obius 
transformations\footnote{diffeomorphisms of $S^{1}$ of the form
$z \mapsto \frac{az+b}{\overline{b}z+\overline{a}}$ with $a,b\in \CC$,
$|a|^2-|b|^2=1$.} of the unit circle $S^1$ satisfying for all 
$I_1,I_2,I
\in \I$ and $\varphi \in \mob$ the following properties.

\begin{itemize}
\item[(i)] {\it Isotony.}
$I_1 \subset I_2 \,\Rightarrow\,
\A(I_1) \subset \A (I_2).$
\item[(ii)] {\it Locality.}
$I_1 \cap I_2 = \emptyset \,\Rightarrow\,
[\A(I_1),\A(I_2)]=0.$
\item[(iii)] {\it Covariance.}
$U(\varphi){\A}(I)U(\varphi)^{-1}={\A}(\varphi(I)).$
\item[(iv)] {\it Positivity of the energy.}
The representation $U$ is of positive energy type:
the conformal Hamiltonian $L_0$, defined by
$U(R_\alpha)=e^{i\alpha L_0}$ where $R_\alpha
\in \mob$ is the anticlockwise rotation by 
an angle of $\alpha$, is positive.

\item[(v)] {\it Existence and uniqueness of the vacuum.}
Up to phase there exists a unique unit vector
$\Omega \in \H_\A$ called the ``vacuum vector''
which is invariant under the action of $U$.

\item[(vi)] {\it Cyclicity of the vacuum.}
${\Omega}$ is cyclic for the von Neumann 
algebra $\A(S^1)\equiv \{\A(I):I\in \I\}''$.
\end{itemize}

There are many known consequences of the above listed axioms. 
We shall recall some of the most important ones referring to  
\cite{FrG,GuLo96} and \cite{FJ} for proves. 
1.\! {\it Reeh-Schlieder property:}
$\Omega$ is a cyclic and separating vector of the algebra $\A(I)$
for every $I \in \I$. 2.\! {\it Bisognano-Wichmann property:}
$U(\delta^I_{2\pi t})=\Delta_I^{it}$ where ${\Delta}_{I}$ 
is the modular operator associated to $\A(I)$ and $\Omega$, and 
$\delta^I$ is the one-parameter group
of M\"obius transformations preserving the interval
$I$ (the dilations associated to $I$) with
parametrization fixed in the beginning
of the next section. 3.\! {\it Haag-duality:} 
$\A(I)'= \A(I^c)$ for every $I \in \I$,
where $I^c$ denotes the interior of the complement set of
$I$ in $S^1$. 4.\! {\it irreducibility}: 
$\A(S^1)={\rm B}(\H_\A)$, where ${\rm B}(\H_\A)$ 
denotes the algebra of all bounded linear operators on $\H_\A$.
5.\! {\it factoriality}: for an $I\in\I$ the algebra 
$\A(I)$ is either just the trivial algebra $\CC \mathbbm 1$ 
(in which case dim$(\H_\A)=1$ and the whole net is trivial) or 
it is a type ${\rm I\!I\!I}_1$ factor
for every $I \in \I$. 6.\! {\it additivity}: if $\S\subset \I$ 
is a covering of the interval $I$ then $\A(I)\subset 
\{\A(J):J\in \S\}''$. Note that by the {\it Bisognano-Wichmann 
property}, since $\mob$ is generated by the dilations (associated to 
different intervals), the representation $U$ is completely determined by 
the local algebras and the vacuum vector via modular structure.

A {\bf locally normal representation} $\pi$ (or for shortness, just simply 
representation) of a M\"obius covariant local net $(\A,U)$ consits of 
a Hilbert space $\H_\pi$ and a normal representation $\pi_I$ of the
von Neumann algebra $\A(I)$ on $\H_\pi$ for each $I \subset \I$ such
that the collection of representations $\{\pi_I : I \in \I \}$ is 
consistent with the {\it isotony}: $I \subset K \Rightarrow \pi_K 
|_{\A(I)} = \pi_I$. It follows easily from the axioms and the 
known properties of local nets listed above that if $I \cap K = 
\emptyset$ then $[\pi_I(\A(I)),\pi_K(\A(K))]=0$, if $\S \subset \I$ 
is covering of $K \in \I$ then $\{\pi_I(\A(I)):I\in\I\}'' \supset 
\pi_K(\A(K))$, if $\S \subset \I$ is a covering of $S^1$ then
$\{\pi_I(\A(I)):I\in\S\}''= \{\pi_I(\A(I)):I\in\I\}''\equiv\pi(\A)$ 
and finally, that for each $I \in \I$ the representation $\pi_I$ is
faithful. The representation $\pi$ is called {\it irreducible}, if 
$\pi(\A)' = \CC \mathbbm 1$.

\subsection{Diffeomorphism covariance} Let $\diff$ be the group of
orientation preserving (smooth) diffeomorphisms of the circle. It is an
infinite dimensional Lie group whose Lie algebra is identified with the 
real topological vector space Vect$(S^1)$ of smooth real vector fields on
$S^1$ with the usual $C^\infty$ topology \cite[Sect. 6]{Milnor} with 
the negative\footnote{The negative sign is ``compulsory'' if we want
the ``abstract'' exponential --- defined for Lie algebras of Lie groups 
--- to be the same as the exponential of vector fields, i.e. the  
diffeomorphism which is the generated flow at time equal $1$.} of the 
usual bracket of vector fields. We shall think of a the vector field 
symbolically written as $f(e^{i\vartheta})\frac{d}{d\vartheta}\in $ 
Vect$(S^1)$ as the corresponding real function $f$. 
We shall use the notation $f'$ (calling it simply the derivative) for 
the function on the circle obtained by derivating with respect to the 
angle: $f'(e^{i\theta})=\frac{d}{d\alpha}f(e^{i\alpha})|_{\alpha=\theta}$.

A strongly continuous projective unitary representation
$V$ of $\diff$ on a Hilbert space $\H$ is a strongly
continuous $\diff \rightarrow \U(\H)/\mathbb T$ homomorphism.
The restriction of $V$ to $\mob \subset \diff$
always lifts to a unique strongly continuous unitary
representation of the universal covering group
$\widetilde{\mob}$ of $\mob$. $V$ is said to be of positive energy type,
if its conformal Hamiltonian $L_0$, defined by the above representation
of $\widetilde{\mob}$ (similarly as in case of a representation of the
group $\mob$) has nonnegative spectrum.

Sometimes for a $\gamma \in \diff$ we shall think of $V(\gamma)$ as a
unitary operator. Although there are more than one way to fix phases,
note that expressions like Ad$(V(\gamma))$ or $V(\gamma) \in \M$ for a von
Neumann algebra $\M \subset {\rm B}(\H)$ are unambiguous. Note also that 
the selfadjoint generator of a one-parameter group of strongly continuous 
{\it projective} unitaries $t \mapsto Z(t)$ is well defined up to a real 
additive constant: there exists a selfadjoint operator $A$ such that 
Ad$(Z(t))=$ Ad$(e^{iAt})$ for all $t \in \RR$, and if $A'$ is 
another selfadjoint with the same property then $A'=A+r\mathbbm 1$
for some $r \in \RR$.

We shall say that $V$ is an extension of the unitary representation $U$ of 
$\mob$ if we can arrange the phases in such a way that 
$V(\varphi)=U(\varphi)$, or without mentioning phases: Ad$(V(\varphi))=$ 
Ad$(U(\varphi))$, for all $\varphi \in \mob$. Note that such an
extension of a positive energy representation of $\mob$ is of positive
energy.
\begin{definition}
\label{diffcov:def}
A M\"obius covariant net $(\A,$U$)$ is said to be {\bf conformal 
(or diffeomorphism)  covariant} if there is a strongly 
continuous projective unitary representation of $\diff$ on $\H_\A$ 
which extends $U$ (and by an abuse of notation we shall 
denote this extension, too, by $U$), and for all
$\gamma \in \diff$ and $I \in \I$ satisfies
\begin{itemize}
\item
$U(\gamma)\A(I)U(\gamma)^* = \A(\gamma(J)),$
\item
$\gamma|_I={\rm{id}}_I \Rightarrow
\rm{Ad}(U(\gamma))|_{\A(I)}=\rm{id}_{\A(I)}$.
\end{itemize}
\end{definition}
Note that as a consequence of {\it Haag duality}, if a diffeomorphism is
localized in the interval $I$ --- i.e. it acts trivially (identically) 
elsewhere --- then, by the second listed property the corresponding 
unitary is also localized in $I$ in the sense that it belongs to $\A(I)$.
Thus by setting
\begin{equation}
\A_U(I)\equiv \{U(\gamma):\gamma\in\diff, 
\gamma|_{I^c} = {\rm id}_{I^c}\}'' \;\;\; (I\in\I)
\end{equation}
we obtain a {\it conformal subnet}: for all $\gamma\in\diff$ 
and $I\in\I$ we have that $\A_U(I)\subset 
\A(I)$ and $U(\gamma) \A_U(I) U(\gamma)^*=\A_U(\gamma(I))$.
The restriction of the subnet $\A_U$ onto the Hilbert space
$\H_{\A_U}\equiv \overline{(\bigvee_{I\in\I} \A_U(I))\Omega}$ is 
again a conformal net, which --- unless $\A$ is trivial ---
by \cite[Theorem A.1]{Carpi2} is isomorphic to a so-called {\it Virasoro
net}. For a representation $\pi$ of $\A$ we set
$\pi(\A_U)\equiv\{\pi_I(\A_U(I)) :I\in\I\}''$.

The smooth function $f: S^1 \rightarrow \RR$, as a vector field on $S^1$,
gives rise to the one-parameter group of diffeomorphisms $t\mapsto$ 
Exp$(tf)$. Hence, up to an additive real constant the selfadjoint 
generator $T(f)$ of $t \mapsto U({\rm Exp(tf)})$ is well defined.
For any real smooth function $f$ on the circle $T(f)$ is essentially 
selfadjoint on the dense set of {\it finite-energy vectors},
i.e. on the algebraic span of the eigenvectors of $L_0$.
By the condition $<\Omega, T(\cdot)\Omega> = 0$ fixing
the additive constant in its definition, $T$ is called the 
{\bf stress-energy tensor} associated to $U$.
It is an operator valued linear functional in the sense that on the
set of finite energy vectors $T(f + \lambda g) = T(f) + \lambda T(g)$
for all $f,g$ real smooth functions on the circle and $\lambda \in \RR$. 
Note that by the second listed condition in Def.\! \ref{diffcov:def} if 
Supp$(f) \subset I$ for a certain $I \in \I$ then $T(f)$ is affiliated
to $\A(I)$.

For a more detailed introduction on the stress-energy tensor see 
for example \cite{CaWe,Carpi2}. The proof of the statements made in 
defining $T$ relies on the so-called Virasoro operators, which
can always be introduced (see the remarks in the beginning of
\cite[Sect. 4]{CaWe} and before \cite[Theorem A.1]{Carpi2}, all
using the results \cite{loke} of Loke), and on the existence of
some ``energy bounds'' (see \cite{GoWa, BS-M}).

In this paper we shall often use nonsmooth functions.
For a function $f\in C(S^1,\RR)$ with Fourier coefficients 
$\hat{f}_n=\frac{1}{2\pi}\int_0^{2\pi} e^{-in\alpha}f(e^{i\alpha})
\,d\alpha$ $(n \in {\mathbb Z})$ we set
\begin{equation}
\|f\|_{\frac{3}{2}}=
\mathop{\sum}_{n \in {\mathbb Z}}|\hat{f}_n|(1+|n|^{\frac{3}{2}}) 
\in \RR^+_0 \cup \{+\infty\}.
\end{equation}
In \cite[Sect. 4]{CaWe} the present author with Carpi proved
that $T$ can be continuously extended to functions with finite 
$\|\cdot\|_{\frac{3}{2}}$ norm as 
\begin{itemize}
\item
if $f,f_n$ $(n \in \NN)$ are real smooth functions on the
circle and $f_n\rightarrow f$ in the $\|\cdot\|_{\frac{3}{2}}$ 
sense then $T(f_n)$ converges to $T(f)$ in the strong resolvent sense,
\item
if $f_n$ $(n \in \NN)$ is a Cauchy sequence of real smooth 
functions with respect to the $\|\cdot\|_{\frac{3}{2}}$ norm then
$T(f_n)$ converges to a selfadjoint operator in the strong resolvent 
sense, which is essentially selfadjoint on the finite energy vectors, 
\item
the real smooth functions form a dense set among the real continuous 
functions with finite $\|\cdot\|_{\frac{3}{2}}$ norm.
\end{itemize}
Thus one can consider $T(f)$ even when $f$ is not smooth but its
$\|\cdot\|_{\frac{3}{2}}$ norm is finite. The following lemma, which was
essentially demonstrated in the proof of \cite[Lemma 5.3]{CaWe} but was
not stated there can be useful in some cases to establish the finiteness
of this norm.
\begin{lemma}
\label{finite1.5norm}
Let $f$ be a (once) differentiable function on the circle. 
Suppose that there exists a finite set of intervals $I_k\in\I$ 
and smooth functions $g_k$ on the circle $(k=1,..,N)$ such that
$\overline{\cup_{k=1}^N I_k}=S^1$ and $f|_{I_k}=g_k|_{I_k}$.
Then $\|f\|_{\frac{3}{2}} < \infty$.
\end{lemma}
\begin{proof}
The conditions mean that $f''$, which is everywhere defined
apart from a finite set of points, has Fourier coefficients 
$\hat{(f'')}_n=-n^2 \hat{f}_n$ and is of bounded variation.
Therefore $|n^2\hat{f}_n| \leq |\frac{{\rm Var}(f'')}{n}|$ 
(see \cite[Sect.\! I.4]{katznelson}), from which the claim 
follows easily.
\end{proof}

In relation with the net $(\A,U)$ the extension to nonsmooth functions 
is still {\it covariant} and {\it local} in the sense of the following 
statement (which again was essentially proved in \cite{CaWe}, but was not 
explicitly stated there).
\begin{proposition}
\label{aff}
Let $\gamma \in \diff$ and $f$ be a real continuous function 
on the circle with both $\|f\|_{\frac{3}{2}} < +\infty$ and 
$\|\gamma_* f\|_{\frac{3}{2}} < +\infty$ where $\gamma_*$
stand for the action of $\gamma$ on vector fields.
Then up to phase factors
$$U(\gamma)\, e^{iT(f)}\, U(\gamma)^*=
e^{iT(\gamma_* f)}.$$
Moreover, if Supp$(f) \subset 
\overline{I}$ where $I \in \I$, then $T(f)$ is 
affiliated to $\A(I)$.
\end{proposition}
\begin{proof}
For the second part of the statement, by the continuity \cite{jors} 
of the net we may assume that Supp$(f)$ is already contained in $I$ 
(and not only in its closure). 
Then according to \cite[Lemma 4.6]{CaWe}, there exists a 
sequence of smooth functions $f_n$ $(n \in \NN)$ converging 
to $f$ in the $\|\cdot\|_{\frac{3}{2}}$ norm whose support 
is contained in $I$. Then, by \cite[Prop. 4.5]{CaWe} $T(f_n)$
converges to $T(f)$ in the strong resolvent sense, and thus
$T(f)$ is affiliated to $\A(I)$ as $T(f_n)$ is affiliated to 
$\A(I)$ for each $n \in \NN$.

The first part of the statement is again obviously true if $f$ 
is smooth, as then $e^{iT(f)}=U({\rm Exp}(f))$ and 
$e^{iT(\gamma_* f)}=$
\begin{equation}
U({\rm Exp}(\gamma_* f)=U(\gamma\circ{\rm 
Exp}(f)\circ\gamma^{-1})=U(\gamma)\, U({\rm Exp}(f))\, U(\gamma)^*.
\end{equation}
Then similarly to the first part, by approximating $f$
with smooth functions and taking limits one can easily 
finish the proof.  
\end{proof}

\section{Proof of the Positivity}

Apart from the subgroup $\mob \subset \diff$, for our argument 
we shall need to use some other important subgroups. For each
positive integer $n$ the group $\mob^{(n)}$ is defined to 
be the subgroup of $\diff$ containing all elements 
$\gamma \in \diff$ for which there exists a M\"obius transformation $\phi
\in \mob$ satisfying
\begin{equation}
\gamma(z)^n = \phi(z^n) \;\; (\forall z \in S^1).
\end{equation}
Thus the group $\mob^{(n)}$ gives a natural 
$n$-covering of $\mob$. This group has been already
considered and successfully used for the analyses of 
conformal nets, see e.g.\! \cite{LoXu}.

In $\mob$, beside the rotations one often considers 
the translations $a\mapsto \tau_a$ and the dilations
$s \mapsto \delta_s$, that are the one-parameter 
groups generated by the vector fields 
$t(z) = \frac{1}{2}-\frac{1}{4}(z+z^{-1})$ and 
$d(z) = \frac{i}{2}(z+z^{-1})$, respectively.
For an $I\in\I$ one may choose a transformation 
$\phi\in\mob$ such that $\phi(S^1_+)=I$ where 
$S^1_\pm = \{z \in S^1: \pm {\rm Im}(z) > 0\}$. The
one-parameter group $s \mapsto \phi\circ \delta_s 
\circ\phi^{-1}$ is independent of $\phi$ (see e.g.\! 
\cite{GLW}) and is called the dilations associated to 
the interval $I$. When no interval is specified, $\delta$ 
always stands for the one associated to $S^1_+$.

By direct calculation $[d,t] = t$ (remember that 
the bracket is the negative of the usual bracket
of vector fields) and thus at the group level we find
\begin{equation}
\label{scaling}
\delta_s \tau_a \delta_{-s} = \tau_{e^s a}
\end{equation}
i.e.\! the dilations ``scale'' the translations.

In $\mob^{(n)}$, just like in $\mob$, one introduces the
one-parameter subgroup of translations $a \mapsto \tau^{(n)}_a$, 
which is defined by the usual procedure of lifting:
it is the unique continuous one-parameter subgroup satisfying
$\tau^{(n)}_a(z)^n = \tau_a(z^n)$. Alternatively, one may define
it directly with its generating vector field 
$t^{(n)}(z)=\frac{1}{2n} - \frac{1}{4n}(z^n+z^{-n})$.
Similarly one introduces the notion of rotations 
$\alpha \mapsto R^{(n)}_\alpha$ and of dilations 
$s \mapsto \delta^{(n)}_s$. Of course the ``$n$-rotations'', 
apart from a rescaling of the parameter, will simply 
coincide with the ``true'' rotations: 
\begin{equation}
\label{n-rotation}
R^{(n)}_\alpha=R_{\alpha/n}. 
\end{equation}

Let us now consider a strongly continuous projective unitary
representation $V^{(n)}$ of $\mob^{(n)}$. The group
$\mob^{(n)}$ is connected and its Lie algebra is isomorphic to 
${\mathfrak{sl}}(2,\RR)$ which is in particular semisimple (in fact 
even simple, but for what follows semisimplicity is enough). Therefore,
as it is well known, the representation $V^{(n)}$ has a unique strongly 
continuous lift $\tilde{V}$ to the universal cover of $\mob^{(n)}$ which is 
a {\it true} representation. As $\mob^{(n)}$ covers $\mob$ in a 
natural way, its universal cover is canonically identified with 
$\widetilde{\mob}$ which is isomorphic to $\widetilde{{\rm SL}(2,\RR)}$.

The following lemma, although contains some well known facts, is 
hereby included for the convenience of the reader. The presented 
proof is an adopted (and slightly modified) version of the proof 
of \cite[Prop. 1]{koester02}.
\begin{lemma}
\label{mob^n}
Let $\tilde{V}$ be a strongly continuous unitary representation 
of $\widetilde{\mob}$ with $H$ and $P$ being the selfadjoint
generator of rotations and translations in $\tilde{V}$, respectively.
Then the following four conditions are equivalent: \\
\indent $1.\;\,H$ is bounded from below, \\
\indent $2.\;\,P$ is bounded from below, \\
\indent $3.\;\,H \geq 0$, \\
\indent $4.\;\,P \geq 0$.
\end{lemma}
\begin{proof}
Let $\tilde{R}$ be the lift of $R$ and set
$P_\pi=\tilde{V}(\tilde{R}_\pi) P \tilde{V}(\tilde{R}_\pi)^*$; 
it is then the selfadjoint generator associated to the one-parameter
group generated by the vector field $t_\pi$ which we get by
rotating $t$ by $\pi$ radian i.e.\! 
$t_\pi(z)=\frac{1}{2}+\frac{1}{4}(z+z^{-1})$. As $P_\pi$ is unitary 
conjugate to $P$ their spectra coincide. Moreover,
as $t+t_\pi =1$ on the G{\aa}rding-domain we have that 
$P+P_\pi=H$ which immediately proves that if $P$ is 
bounded from below or positive then so is $H$. 

As for the rest of the statement, apart from the trivial
indications there remain only to show that if $H$ is bounded
from below then $P$ is positive. Consider the lifted
dilations $s \mapsto \tilde{\delta}_s$. By equation
(\ref{scaling}) one has that $\tilde{V}(\tilde{\delta}_s) 
P \tilde{V}(\tilde{\delta}_s)^* = e^s P$. Moreover, by
direct calculation $[d,t_\pi]=-t_\pi$ so similarly to the case
of translations the dilations also ``scale'' $t_\pi$, but
in the converse direction. Thus $\tilde{V}(\tilde{\delta}_s)
P_\pi \tilde{V}(\tilde{\delta}_s)^* = e^{-s} P_\pi$. So if 
$H\geq r \mathbbm 1$ for some $r$ real (but not
necessarily nonnegative) number then for any vector
$\xi$ in the G{\aa}rding-domain, setting $\eta
= \tilde{V}(\tilde{\delta}_s)^*\xi$ we have that 
\begin{eqnarray}
\label{ineq}
r \|\xi\|^2 = r \|\eta\|^2 &\leq& <\eta,H\eta>\, 
= e^s <\xi,P\xi> + \, e^{-s} <\xi,P_\pi \xi>
\end{eqnarray}
from which, letting $s\rightarrow \infty$ we find that $P \geq 0$.
\end{proof}
If any of the conditions of the above lemma is satisfied, $\tilde{V}$ 
is called a positive energy representation. A projective representation
$V^{(n)}$ of $\mob^{(n)}$ is said to be of positive energy if its
lift to $\widetilde{\mob}$ is of positive energy.

Let us now consider a conformal local net of on the circle 
$(\A,U)$. By equation (\ref{n-rotation}), $U^{(n)}$, 
the restriction of the positive energy representation 
of $U$ of $\diff$ with stress-energy tensor $T$, is a 
positive energy projective representation of $\mob^{(n)}$.
In particular, as $U^{(2)}$ is of positive energy, the selfadjoint 
operator $T(t^{(2)})$ must be bounded from below, since it generates 
the translations for the representation $U^{(2)}$. (Note that $T(t^{(2)})$ 
is bounded from below but not necessary positive: it is not 
{\it the} generator ---  it still generates the same 
projective one-parameter group of unitaries if you add 
a real constant to it.) The function $t^{(2)}(z)=
\frac{1}{4}-\frac{1}{8}(z^2+z^{-2})$ is a nonnegative 
function with two points of zero: $t^{(2)}(\pm 1)=0$. By
direct calculation of the first derivative: $(t^{(2)})'(\pm 1)=0$, 
hence the decomposition
\begin{equation}
\label{decomp}
t^{(2)} = t^{(2)}_+ + t^{(2)}_-
\end{equation}
with the functions $t^{(2)}_\pm$ defined by the condition 
Supp$(t^{(2)}_\pm) = (S^1_\mp)^c$ is a decomposition of $t^{(2)}$ 
into the sum of two (once) differentiable nonnegative functions that
satisfy the conditions of Lemma \ref{finite1.5norm}. Therefore, as it
was explained in the Preliminaries, we can consider the selfadjoint 
operators $T(t^{(2)}_\pm)$.
\begin{proposition}
\label{affiliation}
Let $(\A,U)$ be a conformal net of local algebras 
on the circle with stress-energy tensor $T$. 
Then $T(t^{(2)}_+)$
is affiliated to $\A(S^1_+)$ and $T(t^{(2)}_-)$ 
is affiliated to $\A(S^1_-)$ and so in particular 
they strongly commute. Moreover, the operators
$T(t^{(2)}_\pm)$ are bounded from below.
\end{proposition}
\begin{proof}
Supp$(t^{(2)}_\pm) \subset \overline{S^1_\pm}$ and so by Prop. 
\ref{aff} $T(t^{(2)}_\pm)$ is affiliated to $\A(S^1_\pm)$. 
So if $P_{[a,b]}$ is a nonzero spectral projection of $T(t^{(2)}_+)$
and $Q_{[c,d]}$ is a nonzero spectral projection of $T(t^{(2)}_-)$, then
$P_{[a,b]} \in \A(S^1_+),\; Q_{[a,b]} \in \A(S^1_-)$ and
by the algebraic independence of two commuting factors
(see for example \cite[Theorem 5.5.4]{kadison}) 
$R=P_{[a,b]}Q_{[c,d]} \neq 0$. Of course the 
range of $R$ is invariant for (and included in the domain of) 
$T(t^{(2)}_+) + T(t^{(2)}_-)$ and the restriction of that operator
for this closed subspace is clearly bigger than $a+c$ and smaller 
than $b+d$. Thus
\begin{equation}
{\rm Sp}\left(T(t^{(2)}_+) + T(t^{(2)}_-)\right) \supset
{\rm Sp}(T(t^{(2)}_+)) + {\rm Sp}(T(t^{(2)}_-)).
\end{equation}
To conclude we only need to observe that by equation 
(\ref{decomp}) on the common core of the finite energy 
vectors $T(t^{(2)}_+) + T(t^{(2)}_-) = T(t^{(2)})$,
and as it was said, the latter selfadjoint operator 
is bounded from below.
\end{proof}
\smallskip

{\it Remark.}
The author considered this construction to indicate that if 
$f \geq 0$ then $T(f)$ is bounded from below, which --- 
as it was already mentioned --- by now is a proven fact
(cf.\cite{FewHoll}). The point is the following. If $f$ is 
{\it strictly} positive then, as a vector field on $S^1$, 
it is conjugate to the constant vector field $r$ for some $r>0$. 
Thus, using the transformation rule of $T$ under diffeomorphisms, 
$T(f)$ is conjugate to $T(r)$ plus a constant, and so it is 
bounded from below by the positivity of $T(1)=L_0$. 
The real question is whether the statement remains
true even when $f$ is nonnegative, but not strictly positive
because for example it is {\it local} 
(there is an entire interval on which it is zero). 
One can of course consider a nonnegative function 
as a limit of positive functions, but then one needs to control 
that the lowest point of the spectrum does not go to $-\infty$ while 
taking this limit (which --- in a slightly different manner --- 
has been successfully carried out in the mentioned 
article). However, even without considering limits,
by the above proposition we find nontrivial examples of local 
nonnegative functions $g$ such that $T(g)$
can easily be checked to be bounded from below.
(Take for example $g=t^{(2)}_\pm$ but of course
we may consider conjugates, sums and multiples 
by positive constants to generate even more examples.)
\smallskip

Let us now investigate what we can say about a representation $\pi$
of the conformal net $(\A,U)$. In \cite{D'AnFreKos} it was proved 
that the M\"obius symmetry is continuously implementible in any 
(locally normal) representation $\pi$ by a unique inner projective way. 
By their construction the implementing operators are elements of
$\pi(\A_U)$. Looking at the article, we see that the only structural
properties of the M\"obius subgroup of $\diff$ that the proof uses are 
the following.
\begin{itemize}
\item
There exist three continuous one-parameter groups 
$\Gamma_1, \Gamma_2$ and $\Gamma_3$ in $\mob$, so that every element
$\gamma \in \mob$ can be uniquely written as a product $\gamma=
\Gamma_1(s_1)\Gamma_2(s_2)\Gamma_3(s_3)$ where the parameters
$(s_1,s_2,s_3)$ depend continuously on $\gamma$. (In the 
article $\Gamma_1$ is the translational, $\Gamma_2$ is 
the dilational and $\Gamma_3$ is the rotational subgroup;
which is the so-called Iwasawa decomposition, see \cite{FrG}.)
\item
The Lie algebra of $\mob$ is simple.
\end{itemize}
These two properties hold not only for the subgroup $\mob$, but 
also for $\mob^{(n)}$ where $n$ is any positive integer:  
for all $n$ the Lie algebra of $\mob^{(n)}$ is isomorphic to 
${\mathfrak{sl}}(2,\RR)$, and with the rotations, dilations and 
translations replaced by $n$-rotations, $n$-dilations and 
$n$-translations we still have the required decomposition. Let us 
collect into a proposition what we have thus concluded.
\begin{proposition}
\label{n-cover}
Let $\pi$ be a locally normal representation of the conformal 
local net of von Neumann algebras on the circle $(\A,U)$.
Then for all positive integer $n$ there exists a unique 
strongly continuos projective representation $U^{(n)}_\pi$ of
$\mob^{(n)}$ such that $U^{(n)}_\pi(\mob^{(n)})\subset \pi(\A)$ 
and for all $\gamma \in \mob^{(n)}$ and  $I \in \I$
$$
{\rm Ad}(U^{(n)}(\gamma)) \circ \pi_I = 
{\pi}_{\gamma(I)} \circ {\rm Ad}(U^{(n)}(\gamma)).
$$
Moreover, this unique representation satisfies
$U^{(n)}_\pi(\mob^{(n)})\subset \pi(\A_U)$.
\end{proposition}
We shall now return  to the particular case $n=2$. 
On one hand, the action of the $2$-translation 
$\tau^{(2)}_a$ in the representation $\pi$ can be implemented 
by $U^{(2)}_\pi(\tau^{(2)}_a)$. On the other hand, as
\begin{equation}
U(\tau^{(2)}_a)= e^{iaT(t^{(2)})}=e^{iaT(t^{(2)}_+)} e^{iaT(t^{(2)}_-)}
\end{equation} 
we may try to implement the same action by
$\pi_{S^1_+}(W_+(a)) \pi_{S^1_-}(W_-(a))$, where 
\begin{equation}
W_\pm(a)=e^{iaT(t^{(2)}_\pm)} \in \A_U(S^1_\pm).
\end{equation}
\begin{proposition}
\label{localimplementation}
The unitary operator in $\pi(\A_U)$
$$W_\pi(a):=\pi_{S^1_+}(W_+(a)) \,\pi_{S^1_-}(W_-(a))
=\pi_{S^1_-}(W_-(a))\, \pi_{S^1_+}(W_+(a))$$ 
up to phase coincides with $U^{(2)}_\pi(\tau^{(2)}_a)$.
\end{proposition}
\begin{proof}
It is more or less trivial that the adjoint action of the two unitaries
coincide on both $\pi_{S^1_+}(\A(S^1_+))$ and $\pi_{S^1_-}(\A(S^1_-))$. 
There remain two problems to overcome:
\begin{itemize} 
\item
the algebra generated by these two algebras do not necessarily 
contain $\pi(\A_U)$, so it is not clear why the adjoint action of 
these two unitaries should coincide on the mentioned algebra,
\item
but even if we knew that the actions coincide, the two unitaries, 
although both belonging to $\pi(\A_U)$, for what we know could still 
``differ'' in an inner element. 
\end{itemize}
As for the first problem, consider an open interval $I \subset S^1$ 
such that it contains the point $-1$ and has $1$ in the interior 
of its complement. Note that due to the conditions imposed on $I$, 
the sets $K_\pm:=I \cup S^1_\pm$ are still elements of $\I$.
\begin{lemma} 
If $a \geq 0$ then $W_+(a) \A(I) W_+(a)^* \subset \A(I)$.
\end{lemma}
\begin{proof}[Proof of the Lemma]
Let us take a sequence of nonnegative smooth functions $\phi_n$
$(n=1,2,..)$ on the real line, such that the support of $\phi_n$
is contained in the interval $(-1/n,1/n)$, and its integral is $1$.
Then, exactly as in \cite[Prop. 4.5, Lemma 4.6]{CaWe}, we have that 
$T(\rho_n)$, with $\rho_n$ being the convolution
\begin{equation}
\rho_n(e^{i\theta})\equiv (t^{(2)}_+ * \phi_n)(e^{i\theta})
\equiv \int t^{(2)}_+(e^{i(\theta + \alpha)}) \phi_n(\alpha)\,d\alpha,
\end{equation}
converges to $T(t^{(2)}_+)$ in the strong resolvent sense.

The flow of a vector field given by a nonnegative function on the circle, 
moves all points forward (i.e. anticlockwise). Moreover,
the flow cannot move points from the support of the vector field 
to outside, and leaves invariant all points outside.

The function $\rho_n$ --- being the convolution
of two nonnegative function --- is nonnegative, and 
its support is $S^1_+$ ``plus $1/n$ radius in both direction''. 
Taking in consideration what was said before
it is easy to see that for $n$ large enough 
Exp$(a \rho_n)(I) \subset I$ and consequently
\begin{equation}
{\rm Ad}\left(e^{i a T(\rho_n)}\right)(\A(I))\subset \A(I).
\end{equation}
Then by the convergence in the strong resolvent sense we obtain
what we have claimed. 
\end{proof}
It follows that if $A \in \A(I)$ and $a \geq 0$ then 
\begin{eqnarray} \nonumber
&&\pi_{S^1_+}(W_+(a))\,\pi_I(A)\,\pi_{S^1_+}(W_+(a))^* = \\  
&&\pi_{K_+} (W_+(a)\,A\,W_+(a)^*) = \pi_I (W_+(a)\,A\,W_+(a)^*)
\end{eqnarray}
and thus $\rm{Ad}\left(W_\pi(a)\right)(\pi_I(A))=
\rm{Ad}\left(\pi_{S^1_-}(W_-(a))\,\pi_{S^1_+}(W_+(a)) \right)
(\pi_I(A))=$
\begin{eqnarray}
\nonumber
&=&\rm{Ad}\left(\pi_{S^1_-}(W_-(a))\right)(\pi_I(W_+(a)\,A\,W_+(a)^*))
\\ \nonumber  
&=&\pi_{K_-}(W_-\,(W_+(a)\,A\,W_+(a)^*)\,W_-(a)^*)
\\
&=&\pi_{K_-}U(\tau^{(2)}_a)\,A\,U(\tau^{(2)}_a)^*)
=\rm{Ad}\left(U^{(2)}_\pi(\tau^{(2)}_a)\right)(\pi_I(A))
\end{eqnarray}
where in the last equality we have used the fact that for 
$a \geq 0$ the image of $I$ under the diffeomorphism
$\tau^{(2)}_{a}$ is contained in $K_-$.

We have thus seen that for $a \geq 0$ the adjoint action
of $W_\pi(a)$ and of $U^{(2)}_\pi(\tau^{(2)}_a)$ 
coincide on $\pi_I(A(I))$. Actually, looking at our argument
we can realize that everything remains true if instead of $I$
we begin with an open interval $L$ that contains the point $1$
and has $-1$ in the interior of its complement and we 
exchange the ``+'' and ``-'' subindices. So in fact we have 
proved that for $a \geq 0$ these adjoint actions coincide on both
$\pi_I(A(I))$ and $\pi_{L}(A(L))$ and therefore on the whole algebra
$\pi(\A)$, since we may assume that the union of $I$ and $L$ is the 
whole circle. (The choice of the intervals, apart from the conditions 
listed, was arbitrary.) Of course the equality of the actions, as 
they are obviously one-parameter automorphism groups of $\pi(\A)$,
is true also in case the parameter $a$ is negative. We can now also
confirm that the unitary
\begin{equation}
Z_\pi(a) \equiv W_\pi(a)^* \, U^{(2)}_\pi(\tau^{(2)}_a)
\end{equation}
lies in $\Z(\pi(\A))\cap \pi(\A_U) \subset \Z(\pi(\A_U))$ where 
``$\Z$'' stands for the word ``center''. Thus $a \mapsto Z_\pi(a)$
is a strongly continuous (projective) one-parameter group. 
(It is easy to see that as $Z_\pi$ commutes with both $W_\pi$ 
and $U^{(2)}_\pi$ it is actually a one-parameter group.)

We shall now deal with the second mentioned problem. The $2$-dilations 
$s \mapsto \delta^{(2)}_s$ scale the $2$-translations and preserve the 
intervals $S^1_\pm$. Thus they also scale the functions $t^{(2)}_\pm$ and 
so we get some relations --- both in the vacuum and in the representation 
$\pi$ --- regarding the unitaries implementing the dilations 
and translations and the unitaries that were denoted by $W$ 
with different subindices (see Prop.\! \ref{aff}).
More concretely, with everything meant in the projective sense, 
in the vacuum Hilbert space we have that the adjoint action of 
$U^{(2)}(\delta^{(2)}_s)$ scales the parameter $a$ into 
$e^sa$ in $U^{(2)}(\tau^{(2)}_a)$ and in $W_\pm(a)$ while in 
$\H_\pi$ we have exactly the same scaling of $U^{(2)}_\pi(\tau^{(2)}_a)$ 
and of $\pi_{S^1_\pm}(W_\pm(a))$ by the adjoint action of 
$U^{(2)}_\pi(\delta^{(2)}_s)$. Thus we find that
\begin{equation}
Ad\left(U^{(2)}_\pi(\delta^{(2)}_s)\right)(Z_\pi(a)) = Z_\pi(e^s a),
\end{equation}
but on the other hand of course, as $Z_\pi$ is in the center, the left
hand side should be simply equal to $Z_\pi(a)$. So $Z_\pi(a)=
Z_\pi(e^s a)$  for all values of the parameters $a$
and $s$ which means that $Z_\pi$ is trivial and hence
in the projective sense $W_\pi(a)$ equals to
$U^{(2)}_\pi(\tau_a)$.
\end{proof}
\begin{corollary}
\label{pos:2}
The projective representation $U^{(2)}_\pi$ is of positive energy.
\end{corollary}
\begin{proof}
As the spectrum of the generator of a one-parameter 
unitary group remains unchanged in any normal representation,
by Prop. \ref{affiliation} the selfadjoint generator of the
one-parameter group 
\begin{equation}
a \mapsto \pi_{S^1_+}\left(e^{iaT(t^{(2)}_+)}\right) 
\pi_{S^1_-}\left(e^{iaT(t^{(2)}_-)}\right)
\end{equation}
is bounded from below and by Prop.\! \ref{localimplementation} this 
one-parameter group of unitaries equals to the one-parameter group
$a \mapsto U^{(2)}_\pi(\tau^{(2)}_a)$ in the projective sense. So 
by Lemma \ref{mob^n} the representation $U^{(2)}_\pi$ is of positive
energy.
\end{proof}

Let us now take an arbitrary positive integer $n$. By equation
(\ref{n-rotation}) $R_\alpha \in \mob^{(n)}$ for all
$\alpha \in \RR$, and by definition both $U^{(n)}_\pi(R_\alpha)$
and $U^{(2)}_\pi(R_\alpha)$ implement the same automorphism of
$\pi(\A)$. Since both unitaries are actually elements of 
$\pi(\A_U)\subset \pi(\A)$, they must commute and
\begin{equation}
\label{propotion}
C_{\pi}^{(n)}(\alpha)=(U^{(n)}_\pi(R_\alpha))^* \,\,
U^{(2)}_\pi(R_\alpha)
\end{equation}
is a strongly continuous one-parameter group in 
$\Z(\pi(\A))\cap \pi(\A_U) \subset \Z(\pi(\A_U))$. 

As it was mentioned, by \cite[Theorem A.1]{Carpi2} the
restriction of the subnet $\A_U$ onto $\H_{\A_U}$ 
--- unless $\A$ is trivial, in which case 
dim$(\H_{\A_U})={\rm dim}(\H_\A)=1$  --- is 
isomorphic to a Virasoro net. Thus $\H_{\A_U}$ 
must be separable (even if the full Hilbert space $\H_\A$ 
is not so; recall that we did not assume separability) as
the Hilbert space of a Virasoro net is separable. 

Every von Neumann algebra on a separable Hilbert space 
has a strongly dense separable $C^*$ subalgebra. A von Neumann
algebra generated by a finite number of von Neumann algebras 
with strongly dense separable $C^*$ subalgebras has a
strongly dense $C^*$ subalgebra. Thus considering that for 
an $I\in\I$ the restriction map from $\A_U(I)$ to 
$\A_U(I)|_{\H_{\A_U}}$ is an isomorphism, one can easily 
verify that the von Neumann algebra $\pi(\A_U)$ has a strongly 
dense $C^*$ subalgebra.

We can thus safely consider the direct integral decomposition 
of $\pi(\A_U)$ along its center
\begin{equation}
\label{directint}
\pi(\A) = {\int}_{\!\!\! X}^\oplus \pi(\A)(x) d\mu(x).
\end{equation}
(Even if $\H_\pi$ is not separable, by the mentioned
property of the algebra $\pi(\A_U)$, it can be decomposed
into the direct sum of invariant separable subspaces for 
$\pi(\A_U)$. Then writing the direct integral decomposition 
in each of those subspaces, the rest of the argument can be
carried out without further changes.) For an introduction
on the topic of the direct integrals see for example 
\cite[Chapter 14.]{kadison}.

As it was mentioned the representations $U^{(n)}_\pi$ $(n=1,2,..)$ 
have a unique strongly continuous lift $\tilde{U}^{(n)}_\pi$ to 
$\widetilde{\mob}$ where $\tilde{U}^{(n)}_\pi$ is a true representation. 
Since the group in question is in particular second countable and
locally compact, and all these representations are in $\pi(\A_U)$, 
the decomposition (\ref{directint}) also decomposes these 
representations 
(cf. \cite[Lemma 8.3.1 and Remark 18.7.6]{dixmier}): 
\begin{equation}
\tilde{U}^{(n)}_\pi(\cdot) = {\int}_{\!\!\! X}^\oplus 
\tilde{U}^{(n)}_{\pi}(\cdot)(x) d\mu(x)
\end{equation}
where $\tilde{U}^{(n)}_{\pi}(\widetilde{\mob})(x) \subset \pi(\A)(x)$ 
and $\tilde{U}^{(n)}_{\pi}(\cdot)(x)$ is a strongly continuous
representation for almost every $x \in X$.
\begin{lemma}
\label{posdirectint}
The representation $\tilde{U}^{(n)}_\pi$ is of positive energy if and 
only if $\tilde{U}^{(n)}_{\pi}(\cdot)(x)$ is of positive energy for 
almost every $x \in X$.
\end{lemma}
\begin{proof}
For a $t\mapsto V(t)$ strongly continuous one-parameter group of 
unitaries the positivity of the selfadjoint generator is for 
example equivalent with the fact that $\hat{V}(f)\equiv\int V(t)f(t)dt
= 0$ for a certain smooth, fast decreasing function $f$ whose 
Fourier transform is positive on $\RR^-$ and zero on $\RR^+$.
If $V$ is a direct integral of a measurable family of strongly
continuous one-parameter groups, $V(\cdot)=\int^\oplus_{\! X} 
V(\cdot)(x) d\mu(x)$, then $\hat{V}(f) =\int^\oplus_{\! X}
\hat{V}(f)(x) d\mu(x)$. As $\hat{V}(f)(x) \geq 0$ for almost
every $x\in X$, the operator $\hat{V}(f)$ is zero if and only
if $\hat{V}(f)(x)=0$ for almost every $x\in X$.
\end{proof}
As $C_{\pi}^{(n)}$ is a strongly continuous one-parameter group 
in the center, for almost all $x \in X:\;\tilde{U}^{(n)}_{\pi}
(R_{(\cdot)})(x) = \tilde{U}^{(2)}_{\pi}(R_{(\cdot)})(x)$ 
in the projective sense. 
Therefore, since by Lemma \ref{posdirectint} and 
Corollary \ref{pos:2} in $\tilde{U}^{(2)}_{\pi}(\cdot)(x)$ 
the selfadjoint generator of rotations is positive, also in 
$\tilde{U}^{(n)}_{\pi}(\cdot)(x)$ it must be at least bounded 
from below and hence by Lemma \ref{n-cover} it is actually positive. 
Thus, by using again Lemma \ref{posdirectint} we arrive to the
following result.
\begin{theorem}
\label{mainresult}
Let $\pi$ be a locally normal representation of the conformal local
net of von Neumann algebras on the circle $(\A,U)$. Then the strongly
continuous projective representation $U^{(n)}_\pi$ of $\mob^{(n)}$,
defined by Proposition \ref{n-cover}, is of positive energy for all
positive integers $n$. In particular, the unique continuous inner
implementation of the M\"obius symmetry in the representation $\pi$
is of positive energy.
\end{theorem}

Carpi proved \cite[Prop. 2.1]{Carpi2} that an irreducible 
representation of a Virasoro net $\A_{{\rm Vir},c}$ must be one of those 
that we get by integrating a positive energy unitary representation 
of the Virasoro algebra (corresponding to the same central charge) 
under the condition that the representation is of positive energy. 
Thus by the above theorem we may draw the following conclusion.
\begin{corollary}
An irreducible representation of the local net $\A_{{\rm Vir},c}$ must 
be one of those that we get by integrating a positive energy unitary 
representation of the Virasoro algebra corresponding to the same 
central charge.
\end{corollary}

\noindent
{\bf Acknowledgements.} The author would like to thank Sebastiano Carpi     
and Roberto Longo for useful discussions, for finding some mistakes and   
for calling his attention to the need for more rigor at certain points
(e.g.\! the need for considerations about separability in respect to the
direct integral decomposition).

\end{document}